\documentclass[%
reprint,
 amsmath,amssymb,
 aps,
prd,
superscriptaddress,
nofootinbib,
floatfix
]{revtex4-2}

\usepackage{lipsum}
\usepackage{tikz}
\usepackage{graphicx}
\usepackage{dcolumn}
\usepackage{bm}
\usepackage[per-mode=symbol]{siunitx} 
\usepackage{hyperref}
\usepackage[mathlines]{lineno}
\usepackage{sidecap}
\usepackage{multirow}
\usepackage{placeins}

\sidecaptionvpos{figure}{c}
\usepackage[capitalise]{cleveref}
\Crefname{figure}{Fig.}{Figs.}
\Crefname{table}{Tab.}{Tabs.}
\Crefname{equation}{Eq.}{Eqs.}

\usepackage{amsmath}
\usepackage{xspace}
\usepackage{modiagram}
\usepackage{upgreek}
\usepackage{chemmacros}
\usepackage{amssymb}
\usepackage{colortbl}
\usepackage{ifthen}

\usepackage{etoolbox}

\makeatletter
\pretocmd{\@makecaption}{\nolinenumbers}{}{}
\makeatother
\newboolean{show_comments}
\setboolean{show_comments}{true}

\newcommand{\bvk}[1]{\ifthenelse{\boolean{show_comments}}{\textcolor{orange}{[{\bf BvK}: #1]}}{}}
\newcommand{\ft}[1]{\ifthenelse{\boolean{show_comments}}{\textcolor{purple}{[{\bf FT}: #1]}}{}}

\DeclareSIUnit\clight{\text{\ensuremath{c}}}

\def \delight {DELight\xspace}

\def \lhe {LHe\xspace}
\def\he#1{He$^{#1+}$}

\begin{document}

\title{Signal partitioning in superfluid \ch{^4He}: a Monte Carlo approach}

\author{Francesco Toschi}\email{francesco.toschi@kit.edu}
\affiliation{Institute for Astroparticle Physics, Karlsruhe Institute of Technology, 76131 Karlsruhe, Germany}
\affiliation{Kirchhoff-Institute for Physics, Heidelberg University, 69120 Heidelberg, Germany}

\author{Axel Brunold}
\affiliation{Kirchhoff-Institute for Physics, Heidelberg University, 69120 Heidelberg, Germany}

\author{Lea Burmeister}
\affiliation{Kirchhoff-Institute for Physics, Heidelberg University, 69120 Heidelberg, Germany}

\author{Klaus Eitel}
\affiliation{Institute for Astroparticle Physics, Karlsruhe Institute of Technology, 76131 Karlsruhe, Germany}

\author{Christian Enss}
\affiliation{Kirchhoff-Institute for Physics, Heidelberg University, 69120 Heidelberg, Germany}

\author{Eleanor Fascione}
\affiliation{Kirchhoff-Institute for Physics, Heidelberg University, 69120 Heidelberg, Germany}

\author{Torben Ferber}
\affiliation{Institute of Experimental Particle Physics,  Karlsruhe Institute of Technology, 76131 Karlsruhe, Germany}

\author{Rahel Gabriel}
\affiliation{Institute for Theoretical Particle Physics, Karlsruhe Institute of Technology, 76131 Karlsruhe, Germany}

\author{Lena Hauswald}
\affiliation{Institute of Micro- and Nanoelectronic Systems, Karlsruhe Institute of Technology, 76131 Karlsruhe, Germany}

\author{Felix Kahlhoefer}
\affiliation{Institute for Theoretical Particle Physics, Karlsruhe Institute of Technology, 76131 Karlsruhe, Germany}

\author{Sebastian Kempf}
\affiliation{Institute of Micro- and Nanoelectronic Systems, Karlsruhe Institute of Technology, 76131 Karlsruhe, Germany}
\affiliation{Institute for Data Processing and Electronics, Karlsruhe Institute of Technology, 76131 Karlsruhe, Germany}

\author{Markus Klute}
\affiliation{Institute of Experimental Particle Physics,  Karlsruhe Institute of Technology, 76131 Karlsruhe, Germany}

\author{Belina von Krosigk}
\affiliation{Kirchhoff-Institute for Physics, Heidelberg University, 69120 Heidelberg, Germany}
\affiliation{Institute for Astroparticle Physics, Karlsruhe Institute of Technology, 76131 Karlsruhe, Germany}

\author{Sebastian Lindemann}
\affiliation{Physikalisches Institut, Universität Freiburg, 79104 Freiburg, Germany}

\author{Benedikt Maier}
\affiliation{Department of Physics, Imperial College, London SW7 2AZ, United Kingdom}
\affiliation{Institute of Experimental Particle Physics,  Karlsruhe Institute of Technology, 76131 Karlsruhe, Germany}

\author{Marc Schumann}
\affiliation{Physikalisches Institut, Universität Freiburg, 79104 Freiburg, Germany}

\author{Melih Solmaz}
\affiliation{Kirchhoff-Institute for Physics, Heidelberg University, 69120 Heidelberg, Germany}
\affiliation{Institute for Astroparticle Physics, Karlsruhe Institute of Technology, 76131 Karlsruhe, Germany}

\author{Kathrin Valerius}
\affiliation{Institute for Astroparticle Physics, Karlsruhe Institute of Technology, 76131 Karlsruhe, Germany}

\author{Friedrich Carl Wagner}
\affiliation{Institute of Micro- and Nanoelectronic Systems, Karlsruhe Institute of Technology, 76131 Karlsruhe, Germany}

\collaboration{DELight Collaboration}
\noaffiliation

\date{\today}

\begin{abstract}
Superfluid \ch{^4He} is an ideal candidate for the direct detection of light dark matter via nuclear recoils thanks to its low nuclear mass and the possibility to reach a low detection energy threshold by exploiting the generated quasiparticles.
The design of future detectors based on this target, such as the DELight experiment, requires a proper understanding of the formation and partitioning of the signal for different energy depositions from various sources.
This work presents an overview of the physical processes involved in the energy deposition of recoiling electrons and ions, and describes a Monte Carlo approach to the partitioning of the signal into different channels.
Despite an overall good agreement with existing literature, differences in the region of interest for light dark matter searches below \SI{200}{\eV} are observed. 
\end{abstract}

\maketitle

\section{\label{sec:intro}Introduction}

Superfluid \ch{^4He} (\lhe) is a promising target material for direct detection of light dark matter (LDM) and it has attracted large interest for its use since the early 1990s.
Unlike heavier noble elements like xenon and argon, which have been successfully used in time projection chambers to set strong limits on weakly interacting massive particles (WIMPs) with masses above \SI{1}{\giga\eV\per\clight\squared} \cite{XENON:2023cxc,LZ:2022lsv,PandaX-4T:2021bab,DarkSide:2018kuk}, helium offers unique advantages for detecting sub-GeV LDM particles.
The low nuclear mass of helium maximizes the momentum transfer from less massive LDM particle candidates, while its superfluid phase enables the production and propagation of quasiparticles in the form of phonons and rotons.
The detection of quasiparticles is crucial for achieving an energy threshold below \SI{100}{\eV} which is necessary to detect the tiny recoil induced by an LDM particle.
This is possible using cryogenic calorimeters like transition edge sensors (TESs) or magnetic microcalorimeters (MMCs) \cite{Pretzl2020,CRESST:2022lqw,Krantz:2023vpe}.
Other advantages include the intrinsic purity of the \lhe target due to the impurities freezing out at mK temperatures, and the low cost.

A first \lhe-based detector was already proposed in the 1990s by the HERON collaboration for the detection of solar neutrinos, although the capability to search for LDM was already known \cite{Lanou:1987eq,Adams:1996ge}.
HERON utilized TESs to detect scintillation light and quasiparticles via quantum evaporation, proving the working principle of such detectors and measuring the response of \lhe to $\alpha$ particles and electrons \cite{Bandler:1995bs,Adams:1995mk,Adams:1998}.
Over the last decade the progress of cryogenic refrigeration and calorimetry has opened new possibilities for the use of this target material.
The Direct search Experiment for Light dark matter (\delight) project aims at instrumenting about \SI{10}{\liter} of \lhe with MMC-based wafer calorimeters, reaching an overall threshold of \SI{20}{\eV} in its first phase~\cite{vonKrosigk:2022vnf}.
This will allow DELight to set constraints to the LDM-nucleon scattering cross section $<10^{-39}\,\unit{\cm\squared}$ at a mass of \SI{200}{\mega\eV\per\clight\squared} with an exposure of only \SI{1}{\kg\cdot\day}, while probing the parameter space well below \SI{100}{\mega\eV\per\clight\squared}.
Also the HeRALD experiment aims at using a \lhe target, but instrumented with TES sensors \cite{Hertel:2018aal}. 
The SPICE/HeRALD collaboration has successfully measured the gain from the quantum evaporation channel and proved an energy threshold of \SI{145}{\eV} at 5 sigma, consistent with a potential sub-\unit{\giga\eV\per\clight\squared} DM search \cite{SPICE:2023aqd}.

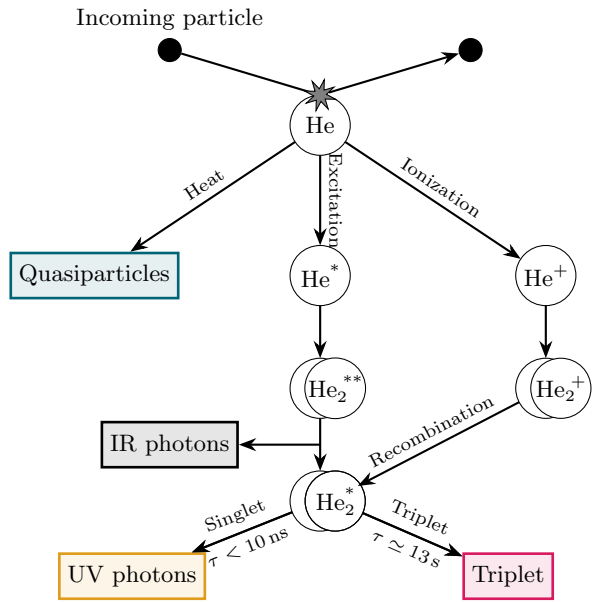
\begin{figure}
    \centering
    \usetikzlibrary{arrows.meta, positioning}
\usepgflibrary{shapes.geometric}
\definecolor{qp_color}{HTML}{006477}
\definecolor{ir_color}{HTML}{000000}
\definecolor{uv_color}{HTML}{DF9B1B}
\definecolor{tr_color}{HTML}{D81B60}

\begin{nolinenumbers}
\begin{tikzpicture}[
  dm/.style={circle, draw, fill=black, minimum size=3mm},
  helium/.style={circle, draw, minimum size=8mm, inner sep=0, outer sep=0},
  dimer/.style={draw, circle, minimum size=8mm, inner sep=0, outer sep=0, fill=white,
  append after command={node [circle, draw, minimum size=8mm, inner sep=0, outer sep=0, fill=white, xshift=2mm] at (\tikzlastnode){#1}}},
  phonon/.style={rectangle, draw=qp_color, fill=qp_color!10, line width=1, minimum size=6mm},
  photon ir/.style={rectangle, draw=ir_color, fill=ir_color!10, line width=1, minimum size=6mm},
  photon uv/.style={rectangle, draw=uv_color, fill=uv_color!10, line width=1, minimum size=6mm},
  triplet/.style={rectangle, draw=tr_color, fill=tr_color!10, line width=1, minimum size=6mm},
  ionization/.style={rectangle, draw, minimum size=6mm},
  arrow/.style={-Stealth, thick},
  particle/.style={rectangle, draw, minimum size=6mm, fill=gray!20}
]

\node[dm, label=above:Incoming particle] (dm_in) at (-2, 1) {};
\node[dm] (dm_out) at (2, 1) {};

\node (helium_top) at (0, 4mm) {};
\node[helium] (helium) at (0, 0) {He};
\draw[arrow] (dm_in) -- (helium_top) -- (dm_out);
\node[star,fill=black, minimum width=0.1cm, star point ratio=2.5, rotate=30, inner sep=0.6mm, star points=8, fill=gray, draw=black] at (helium_top) {};

\node[phonon] (phonon) at (-3, -2) {Quasiparticles};
\draw[arrow] (helium) -- node[midway, above, sloped, font=\scriptsize] {Heat} (phonon);

\node[helium] (helium_excited) at (0, -2) {\ch{He^*}};
\draw[arrow] (helium) -- node[midway, above, sloped, font=\scriptsize] {Excitation} (helium_excited);

\node[helium] (helium_ionized) at (3, -2) {\ch{He^+}};
\draw[arrow] (helium) -- node[midway, above, sloped, font=\scriptsize] {Ionization} (helium_ionized);
\node[dimer={\ch{He2^{+}}}] (excited_excimer_pos) at (3, -3.5) {};
\draw[arrow] (helium_ionized) -- (excited_excimer_pos);

\node[dimer={\ch{He2^{**}}}] (excited_excimer) at (0, -3.5) {};
\draw[arrow] (helium_excited) -- (excited_excimer);

\node[dimer={\ch{He2^*}}] (excimer) at (0, -5) {};
\draw[arrow] (excited_excimer) -- (excimer);

\draw[arrow] (excited_excimer_pos) -- node[midway, above, sloped, font=\scriptsize] {Recombination} ($(excimer) + (0.5, 0.2)$);

\node[photon ir] (ir) at (-2, -4.25) {IR photons};
\draw[arrow] (0, -4.25) -- (ir);

\node[photon uv] (uv) at (-2.5, -6) {UV photons};
\draw[arrow] (excimer) -- node[midway, above, sloped, font=\scriptsize] {Singlet} (uv);
\draw[arrow] (excimer) -- node[midway, below, sloped, font=\scriptsize] {$\tau<10\,\unit{\ns}$} (uv);

\node[helium] (helium_tmp) at (2mm, -5) {};
\node[triplet] (triplet) at (2.5, -6) {Triplet};
\draw[arrow] (helium_tmp) -- node[midway, above, sloped, font=\scriptsize] {Triplet} (triplet);
\draw[arrow] (helium_tmp) -- node[midway, below, sloped, font=\scriptsize] {$\tau\simeq13\,\unit{\s}$} (triplet);

\end{tikzpicture}
\end{nolinenumbers}
    \caption{\label{fig:signal_partition}Schematic representation of signal formation in \lhe.}
\end{figure}

Energy deposition from ionizing radiation in \lhe results not only in ionization and excitation with formation of \ch{He2} excited dimers (excimers), as is common to all noble liquid targets, but also in the production of quasiparticles which freely propagate in \lhe at temperatures below \SI{100}{\milli\K} \cite{Jackle1973}.
The amount of energy that goes into each channel depends on the nature of the recoiling particle, which determines the initial ionization density of the recoiling tracks.
The different signal partition allows for the discrimination between signal-like nuclear recoil (NR) events, which are caused by neutrons or LDM particles, and the background-like electronic recoil (ER) events, caused by electrons and gammas.
When no electric field is applied across the target, as it is the expected case for DELight, the ionization electrons undergo geminate recombination, thus leaving only excitations and quasiparticles.
Quasiparticles have not yet been observed by detectors inside the \lhe because of the high Kapitza resistance \cite{resistance_solid,Kapitza}, hence their detection relies on quantum evaporation.
This consists of the evaporation of single \ch{^4He} atoms into the vacuum due to the quasiparticles reaching the surface, and the subsequent detection of the evaporated atoms via condensation on the surface of helium-film-free sensors.
Additionally, the production of excimers effectively results in three observable signals: infrared photons (IR) from the de-excitation of highly excited excimers, ultraviolet photons (UV) from the prompt ($\tau_\mathrm{s}<10\,\unit{\ns}$ \cite{PhysRevA.67.062716}) decay of excimers in the singlet state, and the delayed triplet signal from the decay of long-lived ($\tau_\mathrm{t}=\left(\num{13(2)}\right)\,\unit{s}$ \cite{PhysRevA.59.200}) triplet states, whose decay is likely induced at the surface \cite{SPICE:2023aqd}.
A diagram of the processes involved in the energy partition into the different channels is given in \cref{fig:signal_partition}.

The yields of the different observable signals (quasiparticles, UV photons, IR photons and triplets) is important information for the development and design of \lhe-based detectors.
Due to the limited availability of measurements, helium lacks a validated common simulation framework, such as the NEST software~\cite{Szydagis_2011} for liquid xenon and argon.
Fundamental work has been done by Ito and Seidel~\cite{Ito:2013cqa}, Guo and McKinsey~\cite{Guo:2013dt}, and Hertel et al. (HeRALD collaboration)~\cite{Hertel:2018aal}.
They estimated the yields starting from the cross section of individual processes involved in the signal formation and assuming a fixed average energy per process and produced quanta.
This paper presents a Monte Carlo approach to a similar cross section-based estimation of the signal partition with the advantages of considering the quantized nature of the signal and the return of information about the correlation of the different signals.
This is of particular importance to optimize the energy reconstruction and the discrimination between ER and NR, especially at low energies.

The paper is organized as follows. 
The physical processes of the interaction of electrons and ions recoiling in \lhe are discussed in \cref{sec:electronic_recoil} and \cref{sec:nuclear_recoil}, respectively.
Particular focus is given to measurements and models used in the Monte Carlo simulation, the details of which are presented in \cref{sec:simulation}.
The results of the simulations are discussed in \cref{sec:results}, together with an overview of the systematic uncertainties associated with the model and a comparison with the available data.
The conclusions are finally summarized in \cref{sec:conclusions}.
The code developed in this work is shortly discussed in Appendix, and is made freely available at~\cite{zenodo_link}.

\section{\label{sec:electronic_recoil}Electronic recoil}
We start with considering the recoil of an electron in \lhe with an initial energy $E$.
The electron could either be a primary particle, e.g., from $\beta$ decay, or the product of a gamma interacting in the target.
The main processes involved in the recoiling of the electron are elastic scattering off a helium atom, and the excitation or ionization of the atom.
The cross section of each process and the energy distribution of the eventual secondaries are discussed in the following.  

\subsection{Elastic scattering}
\begin{figure}[t]
    \centering
    \includegraphics[width=\linewidth]{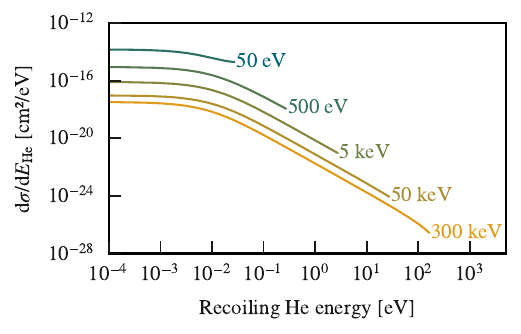}
    \caption{Differential elastic-scattering cross section of electrons off helium atoms for different recoiling electron energies from Ref.~\cite{NIST_el_e}.}
    \label{fig:xsec_el_scattering}
\end{figure}

The differential elastic-scattering cross sections (DCS) of electrons in helium are estimated numerically and tabulated in the NIST database of Ref. \cite{NIST_el_e}.
The DCS is given as a function of the scattering angle $\theta$ in the laboratory frame.
The recoiling energy of the He atom $E_{\mathrm{He}}$ relates to the initial energy of the electron $E$ and the center of mass scattering angle $\phi$ as
\begin{equation}
    E_{\mathrm{He}} = \frac{2r}{\left(r+1\right)^2}\left(1 - \cos\phi\right)E,    
\end{equation}
where $r=M_\mathrm{He} / m_e$, with $M_\mathrm{He}$ the mass of the helium atom and $m_e$ the mass of the electron.
From similar considerations, the scattering angles in the laboratory and center of mass frame are related by ${\cos{\theta}=\left({1+r\cos\phi}\right)\left(r^2 + 2r\cos\phi+1\right)^{-\frac{1}{2}}}$.
Considering that $r\simeq7300\gg1$, we can safely approximate $\cos\theta\simeq\cos\phi$.
The DCS can then be rewritten in terms of the energy of the recoiling helium atom:
\begin{equation}
\frac{\mathop{d\sigma}}{\mathop{dE_\mathrm{He}}} = 2\pi\left|\frac{\mathop{d\cos\theta}}{\mathop{dE_\mathrm{He}}}\right|\frac{\mathop{d\sigma}}{\mathop{d\Omega}} = \pi\frac{\left(M_\mathrm{He} + m_e\right)^2}{EM_\mathrm{He}m_e}\frac{\mathop{d\sigma}}{\mathop{d\Omega}},
\end{equation}
returning the distribution of $E_{\mathrm{He}}$ as a function of the initial electron energy $E$.
The DSC plotted in \cref{fig:xsec_el_scattering} shows that electrons with energies as high as \SI{300}{\keV} can lead up to nuclear recoils of \SI{<200}{\eV}.

Given the DSC, it is possible to estimate the average energy transferred to the recoiling helium atom as:
\begin{equation}
\langle E_\mathrm{He}\rangle =\frac{\int\mathcal{E}\frac{\mathop{d\sigma}}{\mathop{d\mathcal{E}}}\mathop{d\mathcal{E}}}{\int\frac{\mathop{d\sigma}}{\mathop{d\mathcal{E}}}\mathop{d\mathcal{E}}},
\end{equation}
which takes into account both the probability of interaction and the energy transferred to the recoiling helium atom.
The resulting values can be fitted to a power law, returning an average recoiling energy after a single scatter of:
\begin{equation} \label{eq:avg_dE_el_eHe}
    \langle E_\mathrm{He}\rangle \mathrm{[meV]}=5.39\times \left(E\mathrm{[eV]}\right)^{0.21}.
\end{equation}

\subsection{Excitation}\label{sec:e_excitation}
\begin{figure}[t]
    \centering
    \includegraphics[width=\linewidth]{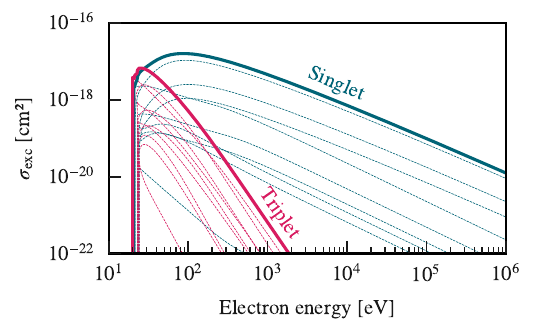}
    \caption{Cross section of the excitation of a ground state helium atom to singlet (blue) and triplet (pink) excited final states due to electron impact. The solid lines represent the total cross sections, while the excitation to the individual excited states are plotted as thin dotted lines.}
    \label{fig:excitation_xsec_el}
\end{figure}

The simulation of the excitation process assumes that all the atoms of the \lhe target are in their $1^1\mathrm{S}$ ground state, since a temperature of \SI{20}{\milli\kelvin} corresponds to a kinetic energy below \SI{2}{\micro\eV}, which is orders of magnitude below the first excitation threshold of \SI{19.82}{\eV} \cite{NIST_he_levels}.
When excited, helium atoms can be found in singlet (S=0) or triplet (S=1) spin states, respectively known as para- and orthohelium.
The cross sections of the excitation processes from the ground state to different excited states are taken from the work by Ralchenko et al.~\cite{RALCHENKO2008603}, where the cross section for each individual excitation is calculated and compared to data.
The cross sections are plotted in \cref{fig:excitation_xsec_el}, where thin dotted lines indicate excitations to specific levels, while the thick solid lines are the total cross section to singlet (blue) or triplet (pink) states.
When the recoiling electron excites a helium atom, it deposits an energy equal to the energy of the excited state $\Delta E_f$.
These values are taken from the NIST database \cite{NIST_he_levels}, and they are reported in \cref{tab:excited_states}.
\begin{table}[h]
\centering
\caption{Excited states of He and their energies sorted in ascending order. The values are from the NIST database \cite{NIST_he_levels}.}
\label{tab:excited_states}
\begin{tabular}{c|c||c|c||c|c}
\textbf{State} & $\Delta E_f$ [eV]  & \textbf{State} & $\Delta E_f$ [eV]  & \textbf{State} & $\Delta E_f$ [eV] \\
\hline
\hline
$2^3\mathrm{S}$ & 19.820 & $3^3\mathrm{P}$ & 23.007 & $4^3\mathrm{P}$ & 23.708 \\
$2^1\mathrm{S}$ & 20.616 & $3^3\mathrm{D}$ & 23.074 & $4^3\mathrm{D}$ & 23.736 \\
$2^3\mathrm{P}$ & 20.964 & $3^1\mathrm{D}$ & 23.074 & $4^1\mathrm{D}$ & 23.736 \\
$2^1\mathrm{P}$ & 21.218 & $3^1\mathrm{P}$ & 23.087 & $4^3\mathrm{F}$ & 23.737 \\
$3^3\mathrm{S}$ & 22.718 & $4^3\mathrm{S}$ & 23.594 & $4^1\mathrm{F}$ & 23.737 \\
$3^1\mathrm{S}$ & 22.920 & $4^1\mathrm{S}$ & 23.674 & $4^1\mathrm{P}$ & 23.742 \\
\end{tabular}
\end{table}

As the helium atoms of the target are excited either to singlet or triplet states, they decay radiatively into the respective first excited state well within \SI{1}{\us}~\cite{PhysRevA.4.1327,Zitnik_2003,Heron_1956,PhysRevLett.103.053002}: $2^3\mathrm{S}$ if triplet, $2^1\mathrm{S}$ if singlet.
This process releases IR radiation whose energy depends on the initial energy level of the excited atom.
The remaining $2^3\mathrm{S}$ and $2^1\mathrm{S}$ excited atoms bind with an helium atom in the ground state to form excited dimers (excimers) in the respective state \ch{He2(a^3$\Sigma$^+_u)} and \ch{He2(A^1$\Sigma$^+_u)}.
This happens with a characteristic time of \SI{15}{\us} for the triplet case~\cite{PhysRevA.10.872}, and \SI{1.6}{\us} for the singlet~\cite{PhysRevA.67.062716}.
The binding energy from the formation of the excimers is released non-radiatively: an average value of \SI{2}{\eV} is considered for both states~\cite{10.1063/1.1673675}.
After the formation of the excimers, the decay differs based on their spin.

Singlet excimers decay within a few nanoseconds via the emission of a UV photon. 
The interatomic distance between the two unbound atoms is roughly \SI{1}{\angstrom}, which corresponds to a potential energy in the unbound system of \SI{3.22}{\eV}, which is released non-radiatively and contributes to the quasiparticle signal~\cite{10.1063/1.4712218,heprops}.
The energy of the emitted UV photons is then the energy of the $2^1\mathrm{S}$ excited states, from which the average dimer binding energy and the unbound potential energy are removed, for a total of around \SI{15.4}{\eV}.

Triplet states have a lifetime of about \SI{13}{\s} and a ballistic propagation speed of \SIrange[range-phrase = --, range-units=single]{2}{4}{\m\per\s}~\cite{2013JLTP..171..207Z}, which means that in a detector of $\sim20\,\unit{cm}$, such as DELight, they come in contact with a surface before decaying.
As this happens, the triplets decay by emitting a UV photon either promptly, if the interface is with a solid, or with a characteristic time of around \SI{5}{\ms}, as reported in Ref.~\cite{SPICE:2023aqd}.
Because of the different time scale, the triplets are considered a different signal with respect to UV photons from singlets, and they carry an energy of \SI{17.82}{\eV}, coming from the difference between the $2^3\mathrm{S}$ excitation energy and the average dimer binding energy.



\subsection{Ionization} \label{sec:e_ionization}
\begin{figure}[t]
    \centering
    \includegraphics[width=\linewidth]{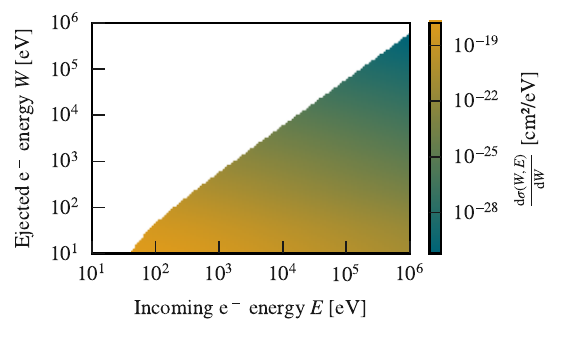}
    \caption{Differential cross section of the electron impact ionization of helium as a function of the incoming and ejected electron energy.}
    \label{fig:diff_xsec_ion_ER}
\end{figure}

\begin{figure}
    \centering 
    \includegraphics[width=\linewidth]{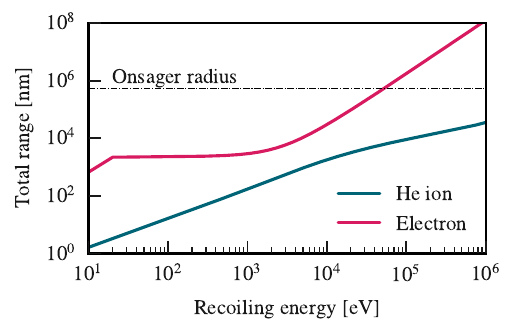}
    \caption{Electron (pink curve) and ion (blue) range in \lhe as a function of the recoiling energy. The dashed black horizontal line indicates the Onsager radius at \SI{20}{\milli\kelvin}. The ion's range is a weighted average of the stopping power for the different charge states.}
    \label{fig:onsager_range}
\end{figure}

The last considered process for the ER interaction is the ionization of the target helium atoms with cross section given in the same reference as the excitation~\cite{RALCHENKO2008603}.
Only single ionization is considered in the following, as double ionization contributes to the total less than \SI{0.5}{\percent}~\cite{Shah_1988}.
The DCS as a function of the ejected electron energy is calculated using the Binary-Encounter-Dipole (BED) model and the values reported in Ref.~\cite{PhysRevA.61.034702}.
The results are plotted in Fig.~\ref{fig:diff_xsec_ion_ER} and they show that even highly energetic primaries most likely produce secondary electrons with energies below \SI{100}{\eV}.
This is orders of magnitude lower than the energy needed for the electron to overcome the electrostatic attraction of the originating ion, which is parameterized by the Onsager radius.
The Onsager radius in a \SI{20}{\milli\kelvin} \lhe bath is \SI{\sim530}{\um}: this is equivalent to the range of a \SI{70}{\keV} electron as shown in \cref{fig:onsager_range}.
Thus, in absence of an electric field, the secondary electrons undergo geminate recombination.

The recombination happens within \SI{0.3}{\nano\s}, the period of time during which the positively charged ion has formed a bound dimer state \ch{He2^+}, releasing non-radiatively the binding energy of \SI{\sim2.4}{\eV}, which contributes to the quasiparticle signal~\cite{10.1063/1.3692800}.
The charged excimer promptly recombines to a final excited and neutral state which is either the triplet \ch{He2(a^3$\Sigma$^+_u)} or singlet \ch{He2(A^1$\Sigma$^+_u)} with equal probability, as determined in Ref.~\cite{phd_adams}.
The recombination radiatively releases an amount of energy which depends on the final spin state.
From this point the behavior is identical to the excitation case discussed at the end of \cref{sec:e_excitation}.

\section{Nuclear recoil}\label{sec:nuclear_recoil}
The processes involved in the recoil of a helium nucleus in the \lhe target, potentially due to an interacting neutron or a LDM particle, are essentially the same discussed for the case of ER. 
Nevertheless, there is an additional degree of complexity in the NR case: while an electron is a fundamental particle with charge $-e$, the recoiling nucleus is part of a complex bound state, the atom, which can be ionized or exchange electrons with the surrounding environment.
We limit our discussion to consider only the neutral and positively ionized cases: \he{0}, \he{1} and \he{2}.
The cross sections for the different processes are considered for each of these charge states, and two additional processes are discussed: charge exchange and Penning quenching.

\subsection{Elastic scattering} \label{sec:nr_elastic}
\begin{figure}[t]
    \centering
    \includegraphics[width=\linewidth]{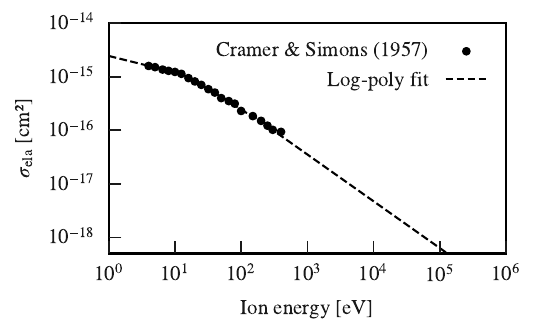}
    \caption{Total elastic cross section for a recoiling ion. The data points are obtained using \he{1}~\cite{10.1063/1.1743506}. The cross section is assumed to be valid for all charge states.}
    \label{fig:nr_tot_xsec_el}
\end{figure}

Measurements by Cramer and Simons~\cite{10.1063/1.1743506} provide the elastic cross section for \he{1} ions.
These values are assumed to be valid for all the charge states and they are empirically fitted using logarithmic polynomials with the degree varying from 1 to 10, and selecting the degree leading to the best reduced $\chi^2$.
This fit procedure is followed for all the measurements presented in the rest of this work.
Measurements and fit are shown in Fig.~\ref{fig:nr_tot_xsec_el}.

The deposited energy in an elastic scattering in the case of target and projectile having the same mass is ${\Delta E = E\sin^2{\theta}}$, where $E$ is the projectile energy and $\theta$ is the scattering angle in the laboratory frame.
We assume a Rutherford-like DSC in the laboratory frame. 
Considering the relationship between the deposited energy and the scattering angle, we obtain:
\begin{equation*}
    \frac{d\sigma}{dE}\propto\frac{1}{\cos\theta}\frac{d\sigma}{d\Omega}\propto\frac{1}{\sin^4\theta}\propto\frac{1}{E^2}.
\end{equation*}
The DSC diverges at low energies: the lower bound is physically constrained by the largest impact parameter achievable, limited by the interatomic distance.
The lower bound $E_\mathrm{min}$ of the Rutherford-like DSC is determined by requiring that the estimated nuclear stopping power agrees with the values from the NIST ASTAR database $\langle E\sigma\rangle$~\cite{astar} and the measurements of $\sigma_\mathrm{el}$ from Cramer and Simons~\cite{10.1063/1.1743506}:
\begin{equation*}
    \frac{\int_{E_\mathrm{min}}^{E}\mathcal{E}\frac{d\sigma}{d\mathcal{E}}d\mathcal{E}}{\int_{E_\mathrm{min}}^{E}\frac{d\sigma}{d\mathcal{E}}d\mathcal{E}} = \frac{EE_\mathrm{min}}{E-E_\mathrm{min}}\ln{\left(\frac{E}{E_\mathrm{min}}\right)} = \frac{\langle E\sigma\rangle\left(E\right)}{\sigma_\mathrm{el}\left(E\right)}.
\end{equation*}
This returns an energy-dependent lower bound ranging from \SI{5}{\eV} up to \SI{20}{\eV}.

\subsection{Excitation} \label{sec:nr_excitation}
\begin{figure*}[t]
  \centering
  \includegraphics[width=\textwidth]{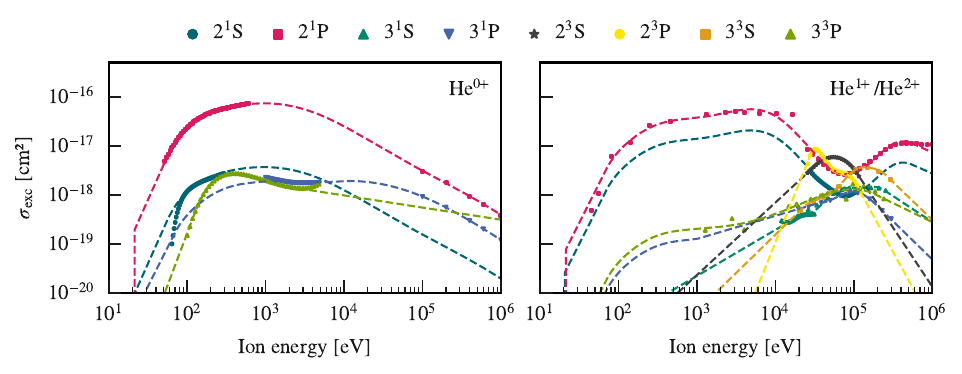}
  \caption{Excitation cross sections for \he{0} (left) and \he{1} (right) for different final excited states. The cross sections for \he{2} are assumed to be identical to \he{1}. Data points are from the ALADDIN database~\cite{aladdin}. The dashed lines indicate the polynomial logarithmic fit discussed in the text.}
  \label{fig:xsec_exc_hehe}
\end{figure*}

An energetic ion propagating in \lhe can excite the surrounding helium atoms from their ground state to higher energy levels.
A large set of measured and theoretical cross sections for these processes are collected in the IAEA ALADDIN database~\cite{aladdin,ornl}.
As only cross sections for \he{0} and \he{1} are available in the database, we make the assumption that the excitation cross sections of \he{2} are identical to the ones of the singly excited counterpart.
A subset of the cross sections for the first excited states is shown in \cref{fig:xsec_exc_hehe}.
Because of the limited energy range of some datasets, great care was taken when extrapolating.
The fit energy range was selected in order to avoid diverging extrapolations at low and high energies, while for some processes the falling edge below \SI{1}{\keV} was determined by scaling the cross section of the 2$^1$P level.
Finally, the 2$^1$S cross section of both charge states was obtained by scaling the 2$^1$P curve: this was motivated by the similar shape and the larger statistics for the 2$^1$P level, especially for the charge states.
The difference in shape below \SI{60}{\eV} for the 2$^1$S cross section of \he{0} has a negligible impact on the final results.
No measurements or calculations of the cross section for the excitation to the 2$^3$S or 2$^3$P states by \he{0} scattering were found in literature, hence they were not included. 
This has a negligible impact on the final results, as it was assessed by using the values from \he{1} for the neutral state.

Previous literature estimated the total excitation cross section by scaling the 2$^1$P results by 1.4 and assuming that the remaining \SI{40}{\percent} was equally split between triplets and singlets~\cite{Ito:2013cqa,Guo:2013dt,Hertel:2018aal}.
This resulted in a ratio of singlets to triplets of 0.86 to 0.14, which is in overall good agreement with the current treatment for energies below \SI{20}{\keV}.

\subsection{Ionization} \label{sec:nr_ionization}
\begin{figure*}[t]
    \centering
    \includegraphics[width=\textwidth]{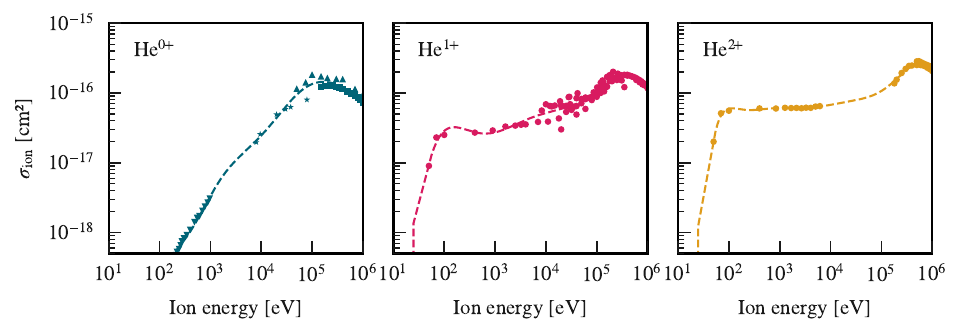}
    \caption{Ionization cross section for \he{0} (left), \he{1} (middle), and \he{2} (right). The cross sections of \he{0} are from: \textcolor[HTML]{006477}{$\blacksquare$}~\cite{PhysRev.178.271}, \textcolor[HTML]{006477}{$\blacktriangle$}~\cite{PhysRevA.76.062710}, \textcolor[HTML]{006477}{$\blacktriangledown$}~\cite{PhysRev.135.A1575}, and \textcolor[HTML]{006477}{$\bigstar$}~\cite{PhysRev.109.385}. The cross sections of the charged states are from~\cite{jaeri_ionization} and references therein. The dashed lines indicate the polynomial logarithmic fit discussed in the text.}
    \label{fig:nr_ionization_xsec}
\end{figure*}

Similarly to excitation, the cross section for the ionization of the target helium depends on the ionization state of the projectile.
The values of the total cross sections for the charged states are taken from the JAERI database~\cite{jaeri_ionization} and for the ground state from various literature~\cite{PhysRev.178.271,PhysRevA.76.062710,PhysRev.135.A1575,PhysRev.109.385}.
The values are plotted in \cref{fig:nr_ionization_xsec}.
The double ionization is neglected as the cross section was measured to be less than \SI{10}{\percent} of the total ionization cross section~\cite{PhysRevA.34.4415}.

The energy distribution of the electron emitted by ionization is described using the Rudd model~\cite{RevModPhys.64.441}, which was originally developed for proton-induced ionization.
The recombination process is identical to the one described for the ER process in \cref{sec:e_ionization}, with the difference that the much larger ionization density of the NR track leads to a columnar recombination (i.e., the electron does not necessarily recombine with the original ion), hence the probability to recombine into a singlet state is \SI{25}{\percent}~\cite{Ito:2013cqa}.

\subsection{Charge exchange}\label{sec:nr_charge_exchange}
\begin{figure}[t!]
    \centering
    \includegraphics[width=\linewidth]{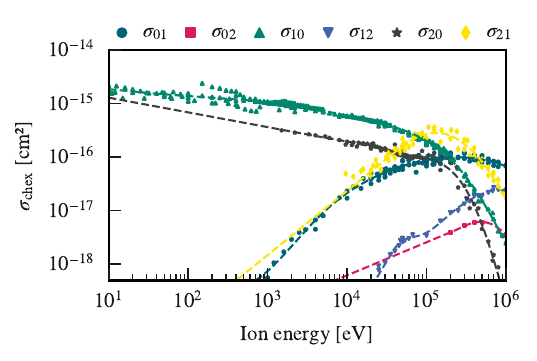}
    \caption{Cross sections of charge exchange processes from the JAERI database~\cite{jaeri_chex}. The notation $\sigma_{if}$ indicates that the recoiling ion has an initial charge state \he{i} and final state \he{f}. The dashed lines indicate the polynomial logarithmic fit discussed in the text.}
    \label{fig:nr_charge_exchange}
\end{figure}

The process of charge exchange is, as the name suggests, the exchange of one or two electrons between the recoiling ion and the surrounding target.
This leads to the change of the charge state of the recoiling ion, together with the emission or capture of one or two electrons.
The cross section of the process depends on the energy of the NR, as well as on the initial and final charge state.
The cross sections are from the JAERI database~\cite{jaeri_chex} and references therein, and they are plotted in \cref{fig:nr_charge_exchange}.

The ionization of the recoiling ion is equivalent to the ionization described in \cref{sec:nr_ionization}, but in the reference frame of the target ion, i.e., the laboratory frame.
The energy distribution of the emitted electrons is hence described by the Rudd model in a boosted frame.
For NR energies up to the MeV scale, the electron's kinetic energy in the laboratory frame $K_e$ is well-approximated using a non-relativistic transformation:
\begin{equation*}
    K_e \overset{v_{\mathrm{CM}} \sim v_{\mathrm{He}}}{\simeq} K_e^\prime + \frac{m_e}{M_\mathrm{He}}K_\mathrm{He}+2\cos\phi\sqrt{\frac{m_e}{M_\mathrm{He}}K_e^\prime K_\mathrm{He}},
\end{equation*}
where $\phi$ is the emission angle in the center of mass frame, $K_\mathrm{He}$ is the kinetic energy of the recoiling helium, and the center of mass speed $v_\mathrm{CM}$ is approximated by the incoming projectile velocity.
The prime indicates the quantity in the center of mass frame.
Although the boost can both increase and decrease the electron energy, the overall effect is to shift the energy distribution to larger values, when a uniform distribution of $\phi$ is assumed.

\begin{figure}
    \centering
    \includegraphics[width=\linewidth]{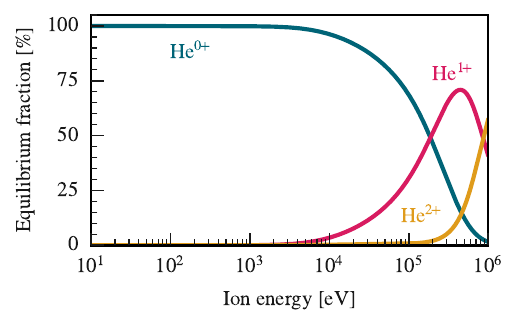}
    \caption{Probability of a helium ion to be found as \he{0} (blue line), \he{1} (pink) or \he{2} (yellow) in equilibrium, i.e., after infinite non-dissipative collisions.}
    \label{fig:ch_equilibrium}
\end{figure}

The cross sections of the different charge exchange processes are used to determine the equilibrium probability of the charge state of a helium ion at a given energy.
In this case, the equilibrium is defined as the charge condition after an infinite number of non-dissipative collisions, hence ensuring no variation of the ion energy.
The results following the calculations of Guo and McKinsey\cite{Guo:2013dt} are shown in \cref{fig:ch_equilibrium}.

\subsection{Penning quenching}\label{sec:penning_quenching}
As an ionizing particle propagates through the medium, it interacts with the surrounding atoms creating a cloud of excited singlet and triplet states from excitation and ionization.
The ionization centers of an ER are on average \SI{850}{\nano\m} apart, while for a NR this distance is around \SI{1.7}{\nano\m}~\cite{Guo:2013dt}.
The very different ionization density implies that, once produced, excimers formed by an NR are densely packed and they can undergo bimolecular (or Penning) quenching~\cite{Ito:2013cqa}:
\begin{align}
\begin{aligned} \label{eq:penning_process}
    \mathrm{He}_2^* + \mathrm{He}_2^* \,&\rightarrow2\mathrm{He} + \mathrm{He_2}^+ + \mathrm{e}^-, \\
    &\rightarrow3\mathrm{He} + \mathrm{He}^+ + \mathrm{e}^-.
\end{aligned}
\end{align}
This process quenches the final observable fraction of UV photons and triplets, in favor of a larger quasiparticle population.

The density of excimers over time ($n_s$ singlets, $n_t$ triplets) is modeled using a set of differential equations as done in~\cite{PhysRevA.85.042718}:
\begin{align*}
\left\{
\begin{aligned}
& \frac{d n_s}{d t}=-\gamma_s\left(k_{s s} n_s^2+k_{s t} n_s n_t\right)-\gamma_t k_{t t} n_t^2-\frac{n_s}{\tau_s} \\
& \frac{d n_t}{d t}=-\gamma_tc_{t t} n_t^2 - \gamma_s \left(c_{s t} n_s n_t +c_{s s} n_s^2\right)-\frac{n_t}{\tau_t}
\end{aligned}
\right.
\end{align*}
where the bimolecular decay coefficients $\gamma_s$ and $\gamma_t$ indicate the decay rate of the excimers via \cref{eq:penning_process}, $\tau_s=10\,\si{\nano\s}$ and $\tau_t=13\,\si{\s}$ are the excimers' lifetimes, and the $k$ and $c$ factors take into account that for each pair of quenched excimers, a new one will be formed after recombination with a singlet-to-triplet ratio of 1:3.

The value used for the bimolecular decay coefficient $\gamma_s$ is taken from Ref.~\cite{PhysRevA.85.042718} and scaled by a factor 9 returning $\gamma_s=\num{2.6e-10}\,\si{\cubic\cm\per\s}$.
This is done because we assume a track of radius \SI{20}{\nm} to estimate the density, while the reference uses a Gaussian with lateral distribution of \SI{60}{\nm}.
The coefficient for the triplet is set to $\gamma_t=0.1\gamma_s$, as it was measured to be around an order of magnitude lower than $\gamma_s$, probably due to the suppressed spin flip required to reach the ground state~\cite{PhysRevA.10.872,Eltsov1998,Kafanov2000}.
As the spin flip argument does not apply when a singlet is involved in the interaction, the $\gamma_s$ coefficient is used to describe the singlet-triplet interaction.

The Penning quenching model allows for the estimation of the energy going into UV photons by considering the fraction of singlets decaying in a period of time $\Delta t$:
\begin{equation} \label{eq:singlet_fraction}
    f_s = \frac{1}{n_s\!\left(0\right)}\int_0^{\Delta t}\frac{n_s}{\tau_s}dt,
\end{equation}
where $n_s\!\left(0\right)$ is the initial singlet density and $\Delta t=50\,\unit{\ns}$.
This is an arbitrary chosen time which is long enough for \SI{>99}{\percent} of the singlets initially produced to decay, but also to include the formation and decay of singlets produced via Penning quenching.
This time is larger than the characteristic diffusion time $t_D=r_0^2/2D_T\simeq5\,\unit{\nano\s}$ with the diffusion coefficient $D_T=\num{4.2e-4}\,\unit{\cm\squared\per\s}$ at a local \lhe temperature of \SI{2}{\kelvin} due to the high quasiparticle density.
Despite this, no diffusion term is included in the model, as the used bimolecular decay coefficient comes from measurement where the term was neglected, hence any contribution from the diffusion is effectively included in the choice of $\gamma_s$~\cite{PhysRevA.85.042718}.
The large uncertainty coming from the choice of $\Delta t$ is later addressed by considering different values.
The triplet contribution is calculated estimating the density of triplets after $\Delta t$.

\begin{figure*}[t]
  \centering
  \includegraphics[width=\textwidth]{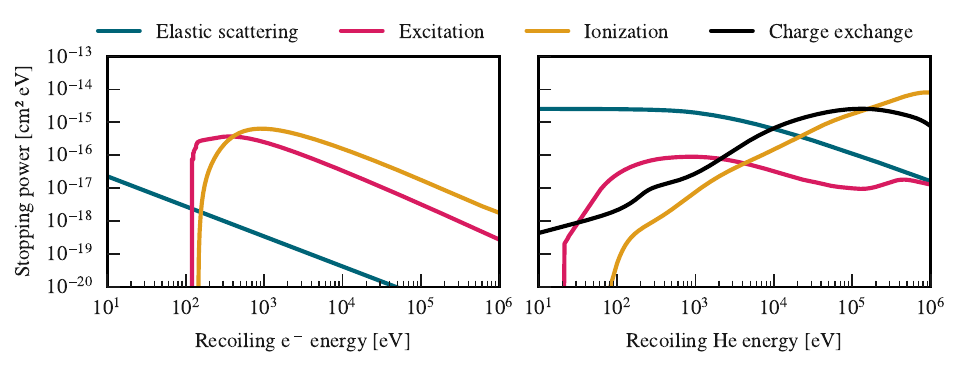}
  \caption[]{Stopping power for different processes in the case of ER (left) and NR (right) events.}
  \label{fig:stopping_power}
\end{figure*}
\section{Simulation}\label{sec:simulation}
The energy partition into the different observable channels for ER and NR is estimated by Monte Carlo simulations.
At the beginning of the simulation, the first process the particle undergoes is sampled using the cross sections of all the physically available processes as respective probability.
This depends on the nature of the recoiling particle (ER or NR), its initial energy $E$, and the charge state, in case the recoiling particle is an ion.
The initial charge state of the recoiling ion is estimated from the equilibrium probability of \cref{fig:ch_equilibrium}.
The fraction of energy going into the different channels for the current process is then stored, and a new process is sampled based on the updated energy and charge state.
This is repeated until the recoiling particle has a kinetic energy $E<19.82\,\unit{\eV}$, as only elastic scattering is possible in this energy region.
The stopping power shown in \cref{fig:stopping_power} is a good indicator of the final signal partitioning, since each process is associated with a given yield of quasiparticles, photons and triplets.
A simplified diagram of the Monte Carlo simulation in the case of ER is shown in the Appendix, together with a brief discussion of the code used for the simulations.

The following discussion uses the physics presented in \cref{sec:electronic_recoil} and \cref{sec:nuclear_recoil} and it summarizes how the energy is partitioned in the different channels for each process.

\subsubsection*{Elastic scattering}
The deposited energy is calculated using \cref{eq:avg_dE_el_eHe} for the ER case and using the Rutherford-like distribution discussed in \cref{sec:nr_elastic} for the NR case.
While for an ER the energy is completely converted into quasiparticles, for an NR this is used as initial energy for the propagation of another ion.

\subsubsection*{Excitation}
The final excited state is sampled from the cross sections by Ralchenko et al.~\cite{RALCHENKO2008603} for the ER case, and from the IAEA ALADDIN database~\cite{aladdin,ornl} for the NR case.
Based on the spin state of the excimer, the energy is shared in different ways among the observable channels.
In case of a singlet excimer: \SI{15.40}{\eV} into UV photons, \SI{5.22}{\eV} into quasiparticles, and the remaining energy into IR.
In case of a triplet excimer: \SI{17.82}{\eV} into triplets, \SI{2.0}{\eV} into quasiparticles, and the remaining energy into IR photons.

\subsubsection*{Ionization}
The ionization process produces a secondary electron which is further propagated, effectively including the effect of secondary electrons in the simulation.
In the case of ER, the initial energy of the emitted electron is sampled from a distribution between 0 and $\left(E - E_\mathrm{ion}\right)/2$ following the BED model described in \cref{sec:e_ionization}.
The upper boundary is due to the indistinguishability of the recoiling and emitted electrons. 
The distribution is binned with bin size \SI{1}{\eV} and a uniformly random energy is chosen within the given bin.
In the case of NR, the electron energy is sampled using the Rudd model of \cref{sec:nr_ionization} which is truncated such that energies with a probability below $10^{-10}$ are neglected.

Electrons and ions are assumed to always recombine to an excimer, either a singlet \ch{He2(A^1$\Sigma$^+_u)} or a triplet \ch{He2(a^3$\Sigma$^+_u)} with a ratio 1:1 for ER and 1:3 for NR.
The energy of \SI{2.4}{\eV} coming from the formation of the \ch{He2^+} dimer is converted into quasiparticles, while the recombination to either one of the neutral excimer states is converted into IR photons: \SI{4.37}{\eV} for triplets and \SI{3.57}{\eV} for singlets.
If the final state is a triplet, the remaining \SI{17.82}{\eV} go into triplet, otherwise \SI{15.40}{\eV} into UV photons and \SI{3.22}{\eV} into quasiparticles.

\subsubsection*{Charge exchange}
Previous approaches to the signal partition in \lhe used averaged cross sections for NR processes weighted by the charge state probability in equilibrium as shown in \cref{fig:ch_equilibrium}.
The Monte Carlo approach uses a well-defined charge state for each step and it includes the possibility for the recoiling ion to change its charge when undergoing a charge exchange process, as discussed in \cref{sec:nr_charge_exchange}.
The charge exchange results in the recoiling ion either capturing or liberating electrons.

In case of the capture of $N$ electrons (where $N$ can be 1 or 2), we assume that the ion loses an energy equivalent to $N$ times the ionization energy $E_\mathrm{ion}=24.59\,\unit{\eV}$~\cite{NIST_he_levels}.
This energy goes into the ionization of a target atom with subsequent recombination, as discussed above for the ionization case.
Although physically the ion does not actually lose kinetic energy as it has potential energy coming from its ionized state, this trick is used to ensure energy conservation when dealing with recombination.
Since electron capture happens only for charged states and they are uncommon for energies below tens of keV (see Fig.~\ref{fig:ch_equilibrium}), such effect is not important when considering the region of interest of LDM recoils.

In case of $N$ ejected electrons, the energy of the $N$ electrons is sampled using the Rudd model and a uniform distribution of the emission angle for its non-relativistic boost: the ion energy is updated accordingly.
No correlation between the $N$ electrons is assumed and they are propagated as independent ERs, exactly like in the case of ionization.

\subsubsection*{Penning quenching}
The processes described so far take part in the pro\-pa\-ga\-tion of the recoiling particle, acting as competing mechanisms with probabilities given by their cross sections.
On the other hand, Penning quenching is applied once the primary particle is left with no kinetic energy and the population of singlet and triplet excimers is final.
This population includes the excimers produced by secondary particles, for which the quenching process is otherwise disabled.
If the number of total excimers is less than two, then Penning quenching is ignored as it cannot physically happen.
Furthermore, the quenching effect is neglected for ERs due to the low ionization density, hence the following discussion applies only to NRs.
 
Given a number of singlets $N_s$ and triplets $N_t$ before quenching, the respective densities are estimated by considering the track to be fully contained in a lateral radius $r_0=\num{20}\,\si{\nm}$ and a length given by the energy-dependent ion range shown in \cref{fig:onsager_range}.
The final number of singlets decaying radiatively, $N_\mathrm{UV}$, is sampled from a Poissonian distribution with an expected value given by the density $n_s\left(\Delta t\right)$ multiplied by the volume, where the choice of $\Delta t$ is discussed in \cref{sec:penning_quenching}.
The energy already partitioned following the excimers' production is then modified considering that $\Delta N_s = N_s - N_\mathrm{UV}$ excimers do not undergo radiative decay.
This means that a total energy $\Delta N_s\cdot \num{15.40}\,\si{\eV}$, which originally went into UV photons, is fully converted into quasiparticles, where the Penning quenching is considered to directly affect only \ch{He2(A^1$\Sigma$^+_u)} and \ch{He2(a^3$\Sigma$^+_u)}.
The number of triplets after quenching is estimated following the discussion of \cref{sec:penning_quenching} and the change, either positive or negative, is equal to $\Delta N_t$.
Similarly to the singlet counterpart, the energy partition is modified by adding a factor $\Delta N_t \cdot \num{17.82} \,\unit{\eV}$ to the triplet component, which is positive if $\Delta N_t>0$ and negative otherwise.
This energy $\Delta N_t \cdot \num{17.82}\,\unit{\eV}$ is eventually added to (or removed from) the quasiparticle channel.

\section{Results} \label{sec:results}
\begin{figure*}[t]
    \centering
    \includegraphics[width=\textwidth]{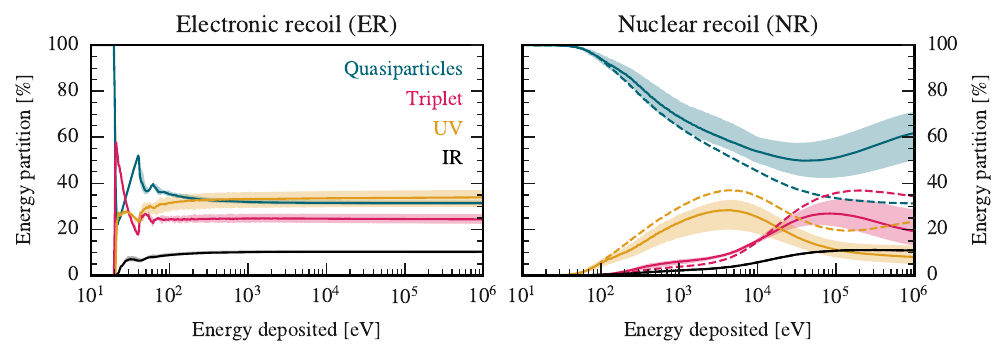}
    \caption[]{Partitioning of the electronic (left) and nuclear (right) recoil energy into the different available signal channels in superfluid \ch{^4He}. The bands indicate the systematic uncertainties, which are discussed in the text. The irregular contours come from the limited statistics used for the simulations. The dashed lines in the right plot indicate the signal partitioning before applying the Penning quenching.}
    \label{fig:energy_partition}
\end{figure*}

The results from the simulations are summarized in \cref{fig:energy_partition}, where the average signal partitioning in the energy range between \SI{10}{\eV} and \SI{1}{\MeV} is shown for ER (left) and NR (right).
Below the energy of the first excited state $2^3\mathrm{S}$ (\SI{19.82}{\eV}) the only process which is allowed is elastic scattering, hence \SI{100}{\percent} of the produced signal consists of quasiparticles.
Although this prevents any possibility to discriminate the nature of the recoil, no gamma-induced ER is physically possible at such low energies, meaning that almost the entirety of the events are NR. 
The behavior of the signal partitioning at the excitation threshold is very different based on the nature of the recoiling particle: while for ERs the triplet channel is promptly populated, elastic scattering is still the dominating energy loss mechanism for NRs (see \cref{fig:stopping_power}) and no clear threshold effect is observed in this case. 
The sharp effect coming from the excitation threshold for ERs is visible also as quantized peaks at around \SI{40}{\eV} and \SI{60}{\eV}.
An electron with these energies can interact with the target via, respectively, two and three excitation processes, while in between these values the remaining energy is converted to quasiparticles via elastic scattering.
As the energy of the recoiling electron increases, the quantized structure is washed out and the signal partitioning of the ER above a few hundred eV is constant with \SI{32}{\percent} of the energy converted into quasiparticle, \SI{33}{\percent} into UV, \SI{24}{\percent} into triplets and the remaining \SI{11}{\percent} into IR.

Unlike for ERs, elastic scattering is the dominant process of energy loss in NRs with energies up to few keV, above which ionizing processes (ionization and charge exchange) are dominant.
Above \SI{5}{\keV} the ionizing processes become the dominant production mechanism of excimers, hence boosting the fraction of singlets and, especially, triplets due to columnar recombination.
This is particularly evident in the signal partitioning before Penning quenching shown in \cref{fig:energy_partition} (right) as dashed lines.
Quenching is important only for energies above \SI{200}{\eV}, when the probability of producing two or more excimers is non-negligible.
The interplay between the decrease of the elastic scattering stopping power and the enhanced quenching effect from the increase of the excimer density results in an overall increase of the quasiparticle signal.
This contribution presents a minimum at around \SI{30}{\keV}, after which it increases at the expense of triplets for higher energies.

\begin{figure}[t]
    \centering
    \includegraphics[width=\linewidth]{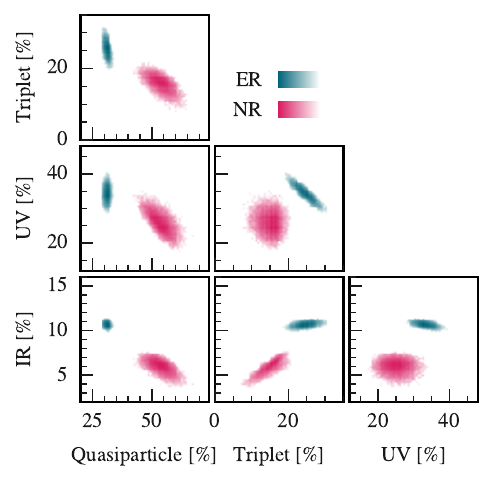}
    \caption{\label{fig:signal_correlation}Correlation among signal contributions in LHe for an energy deposition of \SI{10}{\keV}.
    The blue and pink contours represent ER and NR distributions, respectively.}
\end{figure}

The Monte Carlo approach of the simulation of the \lhe signal partitioning provides insight into the correlation among the different signals at varying recoil energies.
An example of the signal distribution for ER and NR events with an energy of \SI{10}{\keV} is shown in \cref{fig:signal_correlation}.
The ER distribution is characterized by a strong anti-correlation between triplets and UV photons, which can be explained by the fact that excitation and ionization can produce either singlet or triplet excimers, respectively leading to the emission of an UV photon or the propagation of a ballistic triplet.
However, for NRs this correlation is strongly mitigated by the Penning quenching in favor of a stronger anti-correlation between the quasiparticle signal and the UV and triplet signals.

\subsection{Uncertainties}

Together with the average value of the signal fraction, \cref{fig:energy_partition} includes bands representing the uncertainties coming from systematic effects.
In the case of ER, this comes from scaling by \SI{\pm50}{\percent} the cross sections for the excitation, as suggested in the paper by Ralchenko et al.~\cite{RALCHENKO2008603}.
A similar scaling was performed also for the ionization cross section using a \SI{\pm10}{\percent} factor, as coming from the same paper, but the impact is negligible.
The uncertainty of quasiparticle and IR signals is small since a lower excitation cross section does not affect the number of total excimers, but their distribution between singlet and triplet.
This is evident from the much larger uncertainty associated to UV photons and triplets.

The systematic uncertainties in the NR case define broader bands than for its electronic counterpart.
Below \SI{20}{\keV} the dominant systematic uncertainty arises from the \SI{40}{\percent} uncertainty associated to the excitation cross sections~\cite{ornl}.
This is particularly important for the quasiparticle and UV channels, as at these energies excitation and elastic scattering are the dominant energy loss processes.
An increase of the excitation cross section results in a decrease of the quasiparticle contribution in favor of singlet excimers, which are the most likely final states in case of excitation (see \cref{fig:xsec_exc_hehe}).
The uncertainties associated to the cross sections of ionization and charge exchange are respectively \SI{30}{\percent} and \SI{20}{\percent} as discussed in the literature introduced above \cite{jaeri_ionization,PhysRev.178.271,PhysRevA.76.062710,PhysRev.135.A1575,PhysRev.109.385,jaeri_chex}.
Their impact on the signal partitioning is negligible compared to the uncertainties from excitation.

For energies above \SI{20}{\keV} the Penning quenching plays a crucial role in the partitioning of the signal, and the uncertainties associated to it are dominant.
Two main sources of uncertainties are considered for the Penning quenching treatment, in agreement with its discussion in \cref{sec:penning_quenching}: the time period $\Delta t$ of \cref{eq:singlet_fraction} and the bimolecular decay coefficient $\gamma_s$.
The uncertainties associated to these values are estimated by considering the extreme values of \SI{10}{\ns} and \SI{100}{\ns} for $\Delta t$, while the value of $\gamma_s$ was varied between \SI{1e-10}{\cubic\cm\per\s} and \SI{5.5e-10}{\cubic\cm\per\s}, in agreement with values found in literature~\cite{phd_adams}.

\subsection{Comparison to available data}

A modest number of experimental measurements is available for the response of \lhe at energy depositions from electrons and alpha particles.
These are compared to the model developed in this work, showing an overall agreement.

The average energy for the production of an electron-ion pair in \lhe was measured to be around \SI{43}{\eV} both for ER and NR at energies above hundreds of keV \cite{PhysRev.97.1668,Ishida_1992}.
\begin{figure}
    \centering
    \includegraphics[width=\linewidth]{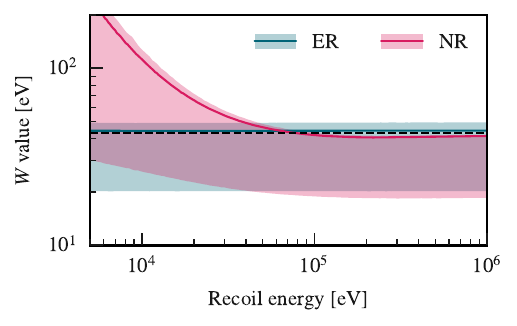}
    \caption{Average energy $W$ for the production of an electron-ion pair in \lhe as coming from simulation of ER (blue) and NR (pink). The black dashed line indicates the measured average energy of \SI{43}{\eV}~\cite{PhysRev.97.1668,Ishida_1992}.}
    \label{fig:w_value}
\end{figure}
The value is properly reproduced both for the ER and NR case, as shown in \cref{fig:w_value}: in the latter case both ionizing processes are considered, ionization and charge exchange.
The large asymmetry in the uncertainties is due to the fact that when decreasing the cross sections of the ionizing processes, their role as dominant processes quickly decreases in favor of non-ionizing excitation.

\begin{figure}
    \centering
    \includegraphics[width=1\linewidth]{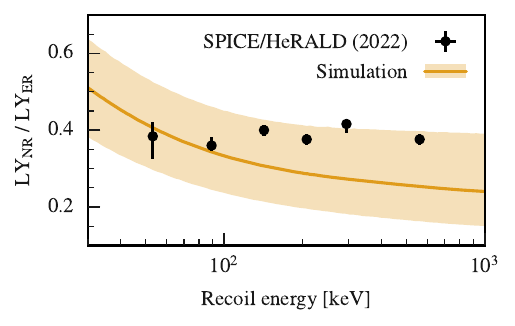}
    \caption{Ratio of NR and ER light yield as measured by the SPICE/HeRALD collaboration~\cite{PhysRevD.105.092005} and from the simulations presented in this work, where only the UV component is considered. Potential sources of differences are discussed in the text.}
    \label{fig:ratio_comparison}
\end{figure}

A complete measurement of the light response of \lhe to ER and NR events in an energy range between few tens of keV to \SI{1}{\MeV} was carried out by the SPICE/HeRALD collaboration~\cite{PhysRevD.105.092005}.
They used a set of six photomultiplier tubes (PMTs) submerged in \lhe at \SI{1.75}{\kelvin} and coupled to a wavelength shifter to be sensitive to the prompt UV light.
The measured ratio between the average ER UV light yield, $\langle\mathrm{LY}_\mathrm{ER}\rangle$, and the NR light yield, $\mathrm{LY}_\mathrm{NR}$, is compared to the simulations.
This is shown in \cref{fig:ratio_comparison}.
Although the simulation agrees with the measurements within the systematic uncertainties, it is possible to observe an overall disagreement at high energies.
The good agreement of the $W$ value indicates that the cause of this difference is the treatment of the Penning effect, which does not affect the intensity of the ionizing process.
The difference could stem from the omission of the diffusion term, which in this work is assumed to be included by the choice of the $\gamma_s$ factor.
As suggested in Ref.~\cite{PhysRevD.105.092005}, dedicated measurements of the UV light yield at different temperatures of \lhe could help to disentangle the effect of diffusion on the signal partition. 


\section{Conclusions} \label{sec:conclusions}
Superfluid \ch{^4He} is a promising target for the direct search of LDM via detection of NR events, thanks to its low nuclear mass and the deposition of energy as quasiparticles that can be detected by instrumenting the active volume with low temperature sensors.
The DELight experiment aims at detecting dark matter by deploying a \SI{10}{\liter} \lhe target at \SI{20}{\milli\kelvin} instrumented using MMCs for an overall threshold of \SI{20}{\eV}~\cite{vonKrosigk:2022vnf}.
The proper knowledge of the signal formation in \lhe is crucial for the design and understanding of the operation of the detector.
This work presented a Monte Carlo-based approach to the estimation of the signal partitioning in \lhe starting from known cross sections of the microphysics processes, building upon previous literature~\cite{Ito:2013cqa,Guo:2013dt,Hertel:2018aal}.
The impact of the uncertainties of the used cross sections on the final partition was assessed, as well as the uncertainties coming from the bimolecular (or Penning) quenching.
The comparison with data returns a good agreement for the $W$ value, while showing a discrepancy of the ratio of the UV yields for NR and ER events.
The latter could be accounted for by a refinement of the Penning quenching model, probably by including a properly tuned diffusion component.

The signal partitioning obtained in this work is in overall good agreement with those already present in literature~\cite{Hertel:2018aal,QUEST-DMC:2023nug}, with notable differences at low energies, between the excitation threshold of \SI{19.82}{\eV} and \SI{200}{\eV}.
The ER partition presents an evident peak structure deriving from the quantization of the excitation energy levels of the target helium atoms, which is mitigated for energies above \SI{100}{\eV}.
This effect is not reproducible when the partition is estimated by averaging the stopping power~\cite{Hertel:2018aal,QUEST-DMC:2023nug}.
The other difference is the absence of a threshold effect in the NR partition: the quasiparticle contribution slowly decreases from \SI{100}{\percent} to approximately \SI{90}{\percent} at around \SI{200}{\eV}, without showing a sharp decrease above the first excitation energy.
This is expected from the fact that the elastic scattering is the dominant contribution to the stopping power in the energy range below \SI{7}{\keV} (see \cref{fig:stopping_power}), while below \SI{100}{\eV} its contribution is larger by more than two orders of magnitude with respect to the excitation contribution.

The observed differences are in the energies of utmost importance for the detection of LDM.
Dedicated calibrations at energies below a few keV are crucial for the resolution of these differences and the proper understanding of \lhe-based detectors for the direct search of LDM.

\begin{acknowledgments}
We thank Prof. Dr. George Seidel for reviewing the first version of this work and his valuable insight which led to the publication of this manuscript.
This work was supported by the Deutsche Forschungsgemeinschaft (DFG) through the Emmy Noether Grant No. 420484612, and by the Alexander von Humboldt Foundation.
\end{acknowledgments}

\appendix*
\section{Monte Carlo code}
\begin{figure}[b]
    \centering
    \resizebox{\linewidth}{!}{\usetikzlibrary{shapes.geometric, arrows, positioning, calc}

\tikzstyle{startstop} = [rectangle, rounded corners, minimum width=3cm, minimum height=1cm, text centered, draw=black, line width=1.5, fill=gray!20]
\tikzstyle{process} = [rectangle, minimum width=2cm, minimum height=1cm, text centered, draw=black]
\tikzstyle{decision} = [diamond, aspect=2, minimum width=3cm, minimum height=1cm, text centered, draw=black]
\tikzstyle{arrow} = [thick,->,>=stealth]
\tikzstyle{io} = [trapezium, trapezium left angle=70, trapezium right angle=110, minimum height=0.7cm, text centered, draw=black]
\tikzstyle{data} = [cylinder, shape border rotate=90, aspect=0.25, minimum height=1.5cm, minimum width=1cm, draw, text centered, fill=gray!20, line width=1.5]

\begin{nolinenumbers}
\begin{tikzpicture}[node distance=2cm]

    \node (input) [startstop] {Input: $E$};
    \node (crosssection) [data, right of=input, xshift=2cm, yshift=-1cm] {Cross sections};
    \node (decision1) [decision, below of=input, yshift=-2mm] {$E > 19.82\,\unit{\eV}$};
    \node (elscattering) [process, below of= decision1, align=center] {Elastic \\ scattering};
    \node (phonons1) [io, below of=elscattering] {QPs};
    \node (excitation) [process, right of= elscattering, xshift=20mm, align=center] {Excitation};
    \node (decision2) [decision, below of=excitation] {Spin state};
    \node (singlet) [io, below of=decision2, yshift=5mm, xshift=-1.5cm, align=center] {IR \\ QPs \\ UV};
    \node (triplet) [io, below of=decision2, yshift=5mm, xshift=1.5cm, align=center] {IR \\ QPs \\ Triplet};
    \node (ionization) [process, right of= excitation, xshift=20mm, align=center] {Ionization};
    \node (ionel) [io, below of=ionization, xshift=-7mm, yshift=8mm] {$e^-$};
    \node (recombination) [process, below of=ionization, xshift=7mm, yshift=-3mm] {Recombination};
    \node (phonons2) [io, below of=recombination, yshift=7 mm] {IR};
    \node (update) [process, below of= decision2, yshift=-1.5cm] {Update $E$};
    \node (elscattering2) [process, at=(elscattering), xshift=-3.5cm, align=center] {Elastic \\ scattering};
    \node (phonons3) [io, below of=elscattering2] {QPs};
    \node (end) [startstop, below of=phonons3] {END};

    \draw [arrow] (input) -- (decision1);
    \coordinate (target) at ($(crosssection|-decision1)$);
    \draw [arrow, dashed] (crosssection) -- (target);
    \draw [arrow] (decision1) -- node[midway, above, sloped] {TRUE} (target) -- ++(0, -1cm) -- ($(elscattering|-target) + (0, -1cm)$) -- (elscattering);
    \draw [arrow] ($(excitation|-target) + (0, -1cm)$) -- (excitation);
    \draw [arrow] (decision1) -- (target) -- ++(0, -1cm) -- ($(ionization|-target) + (0, -1cm)$) -- (ionization);
    \draw [arrow] (elscattering) -- (phonons1);
    \draw [arrow] (excitation) -- (decision2);
    \draw [arrow] (decision2) -- node[midway, left, xshift=-2mm] {S=0} (singlet);
    \draw [arrow] (decision2) -- node[midway, right, xshift=2mm] {S=1} (triplet);
    \draw [arrow] (ionization) -- (ionel);
    \draw [arrow] (ionel) -- ++(2cm, 0) -- ($(ionel|-input) + (2cm, 1cm)$) -- ($(input|-input) + (5mm, 1cm)$) -- (input);
    \draw [arrow] (ionization) -- (recombination);
    \draw [arrow] (recombination) -- (phonons2);
    \draw [arrow] (phonons2) -- ++(-1.5cm, 0) -- ($(excitation) + (15 mm,0)$) -- (excitation);
    \coordinate (target) at ($(elscattering2|-decision1)$);
    \draw [arrow] (decision1) -- node[midway, above, sloped] {FALSE} (target) -- (elscattering2);
    \draw [arrow] (elscattering2) -- (phonons3);
    \draw [arrow] (phonons3) -- (end);
    \draw [arrow] (phonons1) -- ($(phonons1|-update) + (0,10mm)$) -- ($(update|-update) + (0,10mm)$) -- (update);
    \draw [arrow] (phonons2) -- ($(phonons2|-update) + (0,10mm)$) -- ($(update|-update) + (0,10mm)$) -- (update);
    \draw [arrow] (singlet) -- ($(singlet|-update) + (0,10mm)$) -- ($(update|-update) + (0,10mm)$) -- (update);
    \draw [arrow] (triplet) -- ($(triplet|-update) + (0,10mm)$) -- ($(update|-update) + (0,10mm)$) -- (update);    
    \draw [arrow] (update) -- ($(ionel|-update) + (3cm, 0)$) -- ($(ionel|-input) + (3cm, 1.5cm)$) -- ($(input|-input) + (0, 1.5cm)$) -- (input);
\end{tikzpicture}
\end{nolinenumbers}}
    \caption{\label{fig:er_simulation}Flowchart of the ER signal production as discussed in \cref{sec:simulation} and implemented in the \texttt{propagate\_electron} function.}
\end{figure}
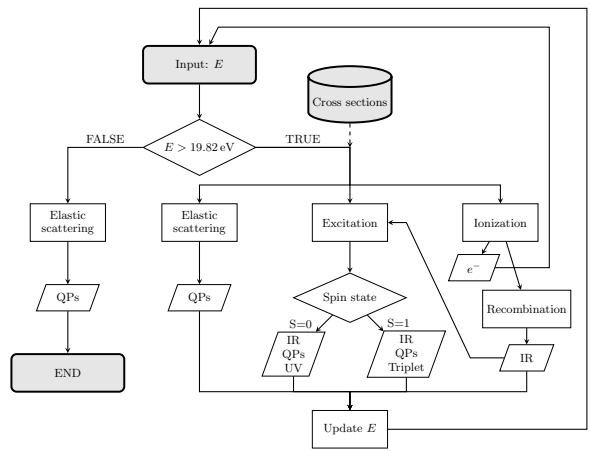

The code developed for the simulations presented in this work is freely available~\cite{zenodo_link}.
The core functions used for the propagation of electrons (ER) and ions (NR) are respectively \texttt{propagate\_electron(E\textsubscript{0})} and \texttt{propagate\_ion(E\textsubscript{0})} from the \texttt{SignalHelium} module, where the argument \texttt{E\textsubscript{0}} is the recoiling energy in eV.
The flowchart of the \texttt{propagate\_electron} function is shown in \cref{fig:er_simulation} and it follows the procedure described in \cref{sec:simulation}.
The flowchart for the ion propagation function is almost identical to what is shown, with the addition of the charge exchange process and the selection and update of the ion charge state.
The cross sections needed for the estimation of the relative intensity of the processes are stored in the \texttt{cross\_sections} directory, while the implementation of the different physical models is done in the utility module \texttt{PhyisicsUtils}. 

The ER propagation function returns three arrays:
\begin{itemize}
    \item \texttt{f} is the energy contribution to each observable channel. They are, respectively, quasiparticles, triplets, UV and IR photons. Their sum is equivalent to the input energy \texttt{E\textsubscript{0}};
    \item \texttt{N} is the number of triplets, UV and IR photons;
    \item \texttt{n} is the number of singlet and triplet excimers produced by ionization and excitation. The order is: singlet from excitation, singlet from ionization, triplet from excitation and triplet from ionization.
\end{itemize}
For reference, the results of \cref{fig:energy_partition} are produced by normalizing \texttt{f} by the input energy.
The NR propagation function returns a similar output, with the difference that \texttt{f} and \texttt{N} contains also respective distribution before the Penning quenching is applied.
The run time is roughly proportional to \texttt{E\textsubscript{0}}$^{3/2}$, where propagating electrons or ions of energy as high as \SI{1}{\MeV} can take up to several minutes of CPU time.

\FloatBarrier
\bibliography{literature}

\end{document}